# FORMATION AND GROWTH OF NANOWIRE

**Sergey P. Fisenko and Felix N. Borovik**

A.V. Luikov Heat and Mass Transfer Institute, National Academy of Science of Belarus,

15, P. Brovka Str., Minsk, 220072, Belarus

E-mail: fsp@hmti.ac.by

**Abstract**

The kinetics of main physical processes controlling the growth of nanowires (NW) via evolution pass " vapor-liquid-solid" is considered. The roles of the thermodynamics and kinetics of cluster nucleation in the initial stage of NW formation are studied. Approximate expressions for NW length are obtained in one-dimensional approximation. The influence of transfer processes in the gas phase on the growth NW is evaluated. The effect of release of the latent heat of phase transition and heat conduction along NW is considered. For carbon NW, grown on Ni catalytic particles, numerical results are obtained.

**Introduction**

Development of the physical fundamentals of the production of nanowires (nanofibres) is one of the actual problems of applied physics. Nanowires (NW) are the very promising products of nanotechnology with broad applications in the near future [1]. Nevertheless there are several unsolved physical problems related to their formation and growth. The aim of our paper is to provide deeper insight into these problems.

It is significant that in many methods of the production of nanoparticles the evolution passway " vapor-liquid-solid" (VLS) is used [1 - 4]. Such the evolution passway means that there are catalytic nanoparticles, which are the key elements of the formation and growth of nanowires.

It is a common knowledge that at the same temperature some properties of objects with radius of tens of nanometers may differ significantly from similar properties of macroobjects. In particular, the melting temperature may drop substantially. We can expect that the diffusion coefficient of the impurity in nanoobjects is close to the diffusion coefficient in liquid phase. Therefore below a catalytic nanoparticle is referred to as nanodroplet. Following [1-3], we assume that a nanodroplet is a melt, which contains two substances $B$ and $A$. The substance $A$ comes to the nanodroplet basically from a gas phase. As is shown below, the crystal substrate substantially decreases the thermodynamic barrier of activation of the formation of impurity nanoparticles. Nevertheless, the substrate is not obligatorily an element of the evolution passway of nanowire formation from nanodroplet. It is well known from different experiments that there are nanowires with top and bottom position of catalytic nanodroplet.

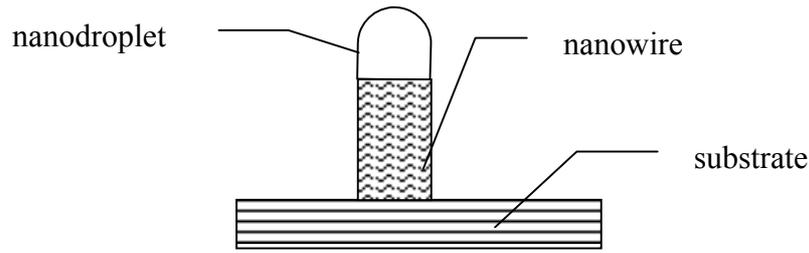

Figure1 Sketch of nanowire on substrate

The scheme of nanowire growth, developed in the framework of the VLS evolution passway, is shown in Fig.1. The physical picture is the following: a supersaturated solution of substance *A* is formed in the nanodroplet of radius R, which is situated on a substrate. Under these conditions there is nucleation on the interfacial border between the nanodroplet and the substrate. The characteristic size of three-dimensional clusters is about 1nm. These clusters grow inside the droplet and merge with one another. After that the nanowire starts to grow already outside the nanodroplet. Obviously, that the diameter of this nanowire is practically the same as that of the nanodroplet.

Thermodynamic reason for the outside growth of nanowire is that this leads to a decreasing of the surface free energy, related with interfacial boundary between the nanodroplet and nanowire. It should be mentioned that the growth inside nanodroplet would lead to a huge increasing of the free energy of the system due to contribution made by deformations of the nanodroplet.

Characteristic size of a nanodroplet cannot be too large. Indeed, it would be difficult to have large supersaturation and an intensive diffusion flow inside. The minimal size of a nanodroplet is discussed below.

The growth rate of a nanowire is determined by diffusion flux inside of a nanodroplet, which depends on the degree supersaturation of the solution inside the nanodroplet and its radius; as well as on the density and structure of nanowire. In turn, the supersaturation of the solution *S* is determined by the mechanisms that maintain the nonequilibrium state.

For a vast majority of experiments such mechanisms are: chemical deposition from a gas phase and, for small partial pressures of vapor, surface diffusion of adatoms. We think that heat processes start to play a substantial role when the length of the nanowire is large enough [5]. In particular, these heat processes are related with the release of the latent heat of phase transition on the nanodroplet – nanowire interface, to vapor condensation on the nanodroplet, and to heat conduction along the nanowire.

We use a continuous approximation to describe of all process below. We recall that the average distance between carbon atoms in a solid phase is about 1.4A. Our minimal spatial scale is about 1nm, therefore the continuous approximation should give reasonable results.

The aim of this paper is further insight into NW growth in the evolution passway VLS, using the theoretical methods of the physical kinetics [6,7]. Molecular dynamics simulation of the growth NW is a tough problem but a very desirable one.

The paper is organized as follows: first we discuss the free energy of formation of three - dimensional clusters and nucleation kinetics. The expression for nucleation rate is obtained. Next, the kinetics of the growth of nanowire is considered, and explicit formulas on the basis of the Stefan approach are derived for a uniform nanowire and a tubular nanowire. Brief discussion of transfer processes in a gas phase is made in free molecular approximation. In the last section we discuss heat problems related to the release of latent heat and substrate temperature. The kinetic results provide insight to the experimental observations. Preliminary results of this research have been published in [4].

**Cluster formation**

The supersaturation S of a solution formed by atoms A in a nanodroplet is

$S = n / n_1(T)$, (1)

where n is the number density of atoms of species A in the solution, $n_l(T)$ is the equilibrium solubility of A at a temperature T. For carbon as an impurity and Ni as the mother phase B, $n_l(873K) \sim 1.9 \cdot 10^{26}$ atoms/m$^3$, and the numerical density of nickel atoms is equal to $9 \cdot 10^{28}$ atoms/m$^3$ [8].

It is well known that the equilibrium solubility of carbon in Ni increases if the temperature of the nanodroplet increases. Thus, we can mention that for the constant numerical density of the impurity the heating of a nanodroplet leads to a decrease in supersaturation of the solution and, correspondingly, to a decrease in the probability of the cluster formation inside the nanodroplet.

Primary clusters are formed due to heterogeneous nucleation near interface between the droplet and the substrate. As is shown below not only properties of the nanodroplet are important but also the properties and temperature of the substrate. For capillary approximation the free energy $\Delta\Phi(g)$ of a one-component cluster of g molecules can be written as follows (for simplicity we accept that clusters have a semispherical shape):

$$\Delta\Phi(g) = -gkT\ln(S) + 2\pi r^2 \sigma_{ab} + \pi r^2 (\sigma_{as} - \sigma_{bs}), \qquad (2)$$

where k is the Boltzmann's constant, r is the cluster radius, $\sigma_{ab}$ is the surface tension between the cluster and phase *B* in the nanodroplet, $\sigma_{as}$ and $\sigma_{bs}$ are respectively the surface tension between the cluster substance (carbon in our case) and the substrate, and between the substrate and phase B. (Fig.2). The last term on right-hand side of expression (2), which takes into account the difference between surface tensions can substantially decrease the free energy of formation of primary clusters.

For many solid substances the values of the surface tension are not known with a good accuracy. This circumstance makes numerical calculations of nucleation rate in a nanodroplet not very accurate one. It should be mentioned only that the value of surface tension is directly proportional to the density difference of substances on interfacial boundary [9].

For very high supersaturations the formation of a cluster at another parts of nanodroplet, which do not contacts with the substrate, is also possible. In this case the formation of clusters should be based on a significantly higher supersaturation in nanodroplet. Indeed for such scenario $\sigma_{bs}=\sigma_{as}=0$. Therefore it is not surprisingly that examples of NW growth with a catalytic nanodroplet on the top are more frequent ones.

Cluster (radius *r*) of *g* atoms, which is formed on the substrate, is determined by the formula

$$r = \left(\frac{3gv_a}{2\pi}\right)^{1/3},$$

where $v_A$ is the volume per atom in the macroscopic volume of *A*.

A critical cluster has *g\** atoms. This value is determined from the condition of the maximum of the free energy of cluster formation $\Delta\Phi(g)$:

$$\frac{\partial \Delta\Phi(g)}{\partial g} = 0. \tag{3}$$

We have the following expression for *g\** (*r\** is the radius of the critical cluster):

$$g^* = \frac{2\pi v_A^2}{3(kT\ln(S))^3}\left[2\sigma_{ab} + (\sigma_{as} - \sigma_{bs})\right]^3. \tag{4}$$

Formula (4) is the generalization of well - known Gibbs's formula [11]. Substituting expression (4) into (2), we find the expression for the free energy of the critical cluster $\Delta\Phi^*$:

$$\Delta\Phi^* = \frac{\pi v_A^2 \left[2\sigma_{ab} + (\sigma_{as} - \sigma_{bs})\right]^3}{3(kT\ln(S))^2}, \qquad (5)$$

or, using the expression for g*,

$$\Delta\Phi^* = \frac{g^* kT \ln(S)}{2}. \qquad (5)'$$

Based on the basic physical idea of nucleation kinetics [10] as the fluctuation transition over the thermodynamic barrier, the approximate formula for the nucleation rate *I* of critical clusters inside a mother droplet is:

$$I = n^2 2\pi D r^* \exp\left[-\Delta\Phi^*/kT\right] \qquad (6)$$

where D is the diffusion coefficient of atoms *A* in the nanodroplet substance B. We neglect the contribution of Zeldovich's factor. The pre-exponential factor in the expression (6) is the product of the normalization constant, which is approximately equal to *n,* and of the one-side diffusion flux on the critical cluster ($2\pi nDr^*$). Let us make numerical estimation. According to (6), for nucleation of carbon in a droplet of Ni (D~$10^{-14}$ m$^2$/s, n~$10^{27}$ atoms/m$^3$, the radius of critical cluster is about 1nm), nucleation rate I is to be about $2 \cdot 10^{23}$ clusters/(m$^3$s) if $\Delta\Phi^*/kT \sim 20$.

Using the expression (4), we can write expression (6) in more detailed form:

$$I = n^2 2\pi D \frac{v_A \left[2\sigma_{ab} + (\sigma_{as} - \sigma_{bs})\right]^3}{kT\ln(S)} \exp\left[-\Delta\Phi^*/kT\right] \qquad (7)$$

For the beginning of the process of NW formation we must have quite a number of clusters formed in the nanodroplet in a short time [fig. 2]. The characteristic time τ for appearance of a new phase cluster in the surface layer of the droplet of radius R can be estimated as [10, 11]

$$\tau \sim R^2 d/I, \qquad (8)$$

where the characteristic height of the layer $d \approx 1$ nm. It follows from the expression given above that for a small nanodroplet the characteristic time for the appearance of clusters is large enough despite of the high nucleation rate. On the other hand $\tau$ is the characteristic time for the overcoming of the nucleation thermodynamic barrier [12, 13]; for temperature about 1000K in a solid $\tau$ typically is about 0.01s. Let us make numerical estimation of the nucleation rate. If $\tau$ is about one millisecond then for a nanodroplet of radius 30 nm, Eq. (8) yields that the nucleation rate is larger than $10^{27}$ clusters/(m$^3$s).

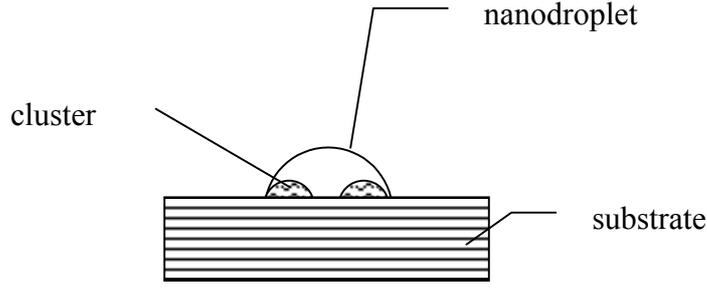

Figure 2 Clusters formation in catalytic nanodroplet

Additionally it is worthwhile to make some geometrical considerations. If the radius of nanodroplet is about 30 nm, we need near 900 critical clusters in order to cover its base. With regards for cluster growth and coalescence we, probably, need two hundred clusters for covering.

After the end of the nucleation time needed for the doubling of the cluster radius is the time for formation of inoculum of NW. This time is about several seconds; it is about $3 \cdot 10^{-18}$ $n_s/Dn$ [14], where $n_s$ is the number density of NW.

**Kinetics of the growth of nanowires**

First, let us consider the growth of a nanowire at isothermal approximation. The coalescence of growing clusters is a fast enough process. It should be noted that the porous structure in the beginning of NW is the possible result from the coalescence of clusters. Experimental verification of this hypothesis is quite simple but important for understanding of the kinetic mechanisms of NW growth. The porous structure in the beginning of NW substantially decreases the total heat conductivity along the nanowire.

After coalescence supersaturation drops in the nanodroplet and the stage of nanowire growth begins with a relatively low supersaturation.

It is worth to emphasize the following conclusion, which follows from our theoretical results: the minimal diameter of a nanowire cannot be smaller than the critical cluster diameter, which is about 1nm.

In one-dimensional approximation for nanowire growth the Stefan condition on the interfacial surface of the nanowire and the nanodroplet is:

$$-D\nabla n = v n_s, \qquad (9)$$

where $v$ is the velocity of motion of the interfacial boundary relative to the substrate. We consider that on the interfacial boundary $n$ is equal to equilibrium number density of A in a nanodroplet.

For the length L of a NW we have the equation:

$$\frac{dL}{dt} = v \qquad (10)$$

Replacing the gradient term in (9) by approximate estimation [14]

$$-\nabla n \approx \frac{n - n_1(T)}{R},$$

we obtain the equation

$$\frac{dL}{dt} = D(T)\frac{n - n_1(T)}{Rn_s}. \tag{11}$$

For constant supersaturation of impurity in the solution S, Eq. (11) can easily be integrated. Thus, we have the important qualitative result

$$L \sim \frac{Dt}{Rn_s}, \tag{12}$$

that is the NW length is directly proportional to time and inversely proportional to the nanodroplet radius. This conclusion was also theoretically obtained in [1- 4] and for a carbon whisker, growing from a Ni particle, was experimentally confirmed in [8].

Let us make numerical estimations of the growth rate of carbon nanowire using expressions (11-12). If $R = 10$nm, typical value of diffusion coefficient is about $10^{-13}$m$^2$/s for a temperature 1000 K, and if the ratio n/n$_s \sim 10^{-1}$, then we have v $\sim 10^{-6}$ m/s.

The formation of tubular nanowires is the subject of special interest. Let us denote the internal radius of a tubular NW by $R_i$, and let the external radius coincide with the radius of the nanodroplet R. Assuming that the full flux of impurity through the nanodroplet provides material for the growth of a tubular nanowire, we have the rate equation for the length of the tubular nanowire L:

$$\frac{dL}{dt} = \frac{DRn_1(T)}{(R^2 - R_i^2)}\frac{(S-1)}{n_s}.$$

Thus, for the same supersaturation the growth rate of the tubular nanowire is higher for small differences of internal and external radii. Generally a tubular NW grows faster than a uniform one.

It should be noted that the characteristic time $\tau_s$ for reaching the steady state profiles of $A$, i.e. the so- called the induction time can be obtained by means of standard estimation for diffusion processes [7]:

$$\tau_s \sim R^2/(\pi^2 D),$$

and, for the nanodroplet of radius 10 nm, $\tau_s$ is about $10^{-4}$ s if diffusion coefficient is about $D \sim 10^{-13}$ m$^2$/s.

The physical picture developed above permits to obtain additional important estimations. Indeed, the minimal radius of nanowires and, correspondingly, the radius of a nanodroplet, is equal to:

$$R = r*.$$

In other words, R must be larger than 1nm. Therefore the minimal height of a nanodropet must be larger than r* too. For tubular nanowires we have a similar estimation for minimal thickness of wall:

$$R - R_i \simeq 2r*,$$

that is the minimal wall thickness is about 2 nm. Therefore for tubular nanowires the minimal radius of nanodroplet should be larger than 3r* (~3 nm). Analysis of experimental data [14,15] confirms our estimations. Obviously formation of a tubular nanowire is possible only in the case where the nanodroplets on substrate have nonuniform height and the maximal height is at the center of the nanodroplet. We can assume that the minimal heights of the nanodroplet should be substantially larger than 2r*, or approximately larger than 2 nm.

Following the developed physical picture of the kinetics of the formation of nanowires, we expect that if the characteristic size of nanodroplet is about 1nm, then the formation of three-dimensional critical cluster, similar to described above one, is practically impossible. But formation of two-dimensional critical clusters is possible. Observations of experimental situations, presented in [14,15], confirm this point of view. One-dimensional clusters give start to the growth of single wall nanotubes. Two –dimensional clusters give possibility to growth to multi -wall nanotubes. Observations of experimental situations, presented in [15], confirm this point of view. The thermodynamics of the formation of two – dimensional clusters has been considered in [14,16].

**Transfer process in gas phase**

Let us consider the transfer process in a gas phase. Clearly this can be done in a free molecular regime. Let us to denote by $J$ the specific flux, which delivers molecules of species $A$ from gas phase on to the nanodroplet surface. For a steady-state regime we then have the equality, representing the conservation law:

$$vn_s = 2J. \qquad (13)$$

Using expression (9) and the estimation of the gradient, we have the relationship between $n$ and $J$:

$$n = \frac{2RJ}{D} + n_1. \qquad (14)$$

It is worthy to emphasize that from (14) we have estimation of the value of the flux $J$:

$$J \sim \frac{Dn_1}{2R} \qquad (15)$$

As a rule, the flux *J* is directly proportional to the partial pressure of the vapor, which carries A in gas phase. From the kinetic theory of gases we have a more exact expression [6],

$$J \sim \frac{P}{\sqrt{mT}},$$

where *m* is the mass of the vapor molecule, which delivery impurity (carbon) to the nanodroplet; some experimental data on such vapors are presented in [17]. For all experimental setups the mass transfer between the nanodroplet and the gas phase occurs at the free molecular regime. Thus, if we increase the partial pressure of vapor we increase the rate of NW growth. If the flux from the gas phase is not sufficient then NW grows with a not constant rate. If the flux from the gas phase exceeds the steady – state one, the additional formation of clusters and nanowires at different parts of a nanodroplet is possible. Such situations have been observed in experiments [18].

## Heating of nanodroplet

Let us make estimations of the heat problems taking place during nanowire growth. Some estimations have been made in [19, 20]. It can be shown that we can neglect the heat transfer between the droplet and the gas phase in comparison with heat transfer through the nanowire. Let us estimate the heat effects of the stage of NW growth. For the temperature profile the characteristic time is much smaller than that of NW growth therefore a quasi - stationary approximation can be used.

On the moving interfacial boundary the law of energy conservation has the form of the Stefan condition [6]:

$$-\lambda \nabla T = v n_s U, \tag{16}$$

where U is the latent heat of phase transition per atom [21], λ is the heat conductivity of the nanowire, where the temperature gradient is reckoned along NW. Replacing the gradient by the expression

$$-\nabla T \approx \frac{T_d - T_s}{L},$$

where $T_d$ and $T_s$ are respectively the temperatures of the nanodroplet and substrate. Substituting the approximate expression for the temperature gradient into equation (16) we have the formula for the quasi steady-state temperature of nanodroplet:

$$T_d(t) \approx T_s + \frac{L(t)DU}{R\lambda} n_s. \qquad (17)$$

It follows from expression (17) that the longer NW, the higher the nanodroplet temperature. Even more, the higher the growth rate, the larger the temperature. Numerical estimation for carbon NW, growing from Ni nanodroplet, shows that for the NW length one micron the temperature difference between the substrate and the nanodroplet does not exceed ten degree (λ ~ 0.1 Wt/(m K), v~$10^{-5}$ m/s, U~5000 k). According to (17), for a 5 micron NW the temperature difference is about 100 K, see also [5]. To obtain more exact results we have to solve jointly the diffusion and heat problems. Such investigation is being carried at present time. Nevertheless it is obvious already that heat effects lead to deviation of the growth law from a linear one.

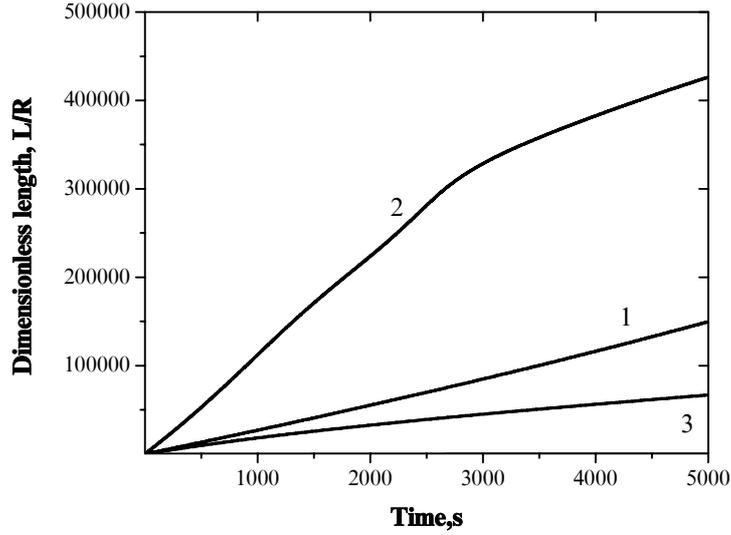

Figure 3. Nanowire length versus time
n =$10^{27}$ atoms/m$^3$, λ=0.2W/Km;
Curve 1-substrate temperature T=1000K, curve 2 - substrate temperature T=1100K, curve 3 - substrate temperature T= 1200K.

The growth of carbon nanowires at different substrate (Ni) temperatures is displayed in Fig. 3. We use dimensionless length of NW *L/R*. For fast growing and quite long nanowire we see deviation from linear law of growth. If we increase the temperature of the substrate, then due to the exponential dependence of diffusion coefficient, we may expect a substantial increase in the rate of growth. But for a constant number density of carbon increasing substrate temperature means decreasing of supersaturation *S,* therefore the curve 2 is below the curve 1. For n=$10^{27}$ atoms/m$^3$ in Ni, the values of supersaturations S versus different temperatures of the substrate are shown in Table 1.

Table. For given number density carbon supersaturation versus substrate (Ni) temperature.

| $T_s$ (K) | 1000 | 1100 | 1200 |
|---|---|---|---|
| S | 2.78 | 1.6 | 1.018 |

As is shown in Fig.3, the deviation from the linear law of growth is based on decrease in supersaturation; even an increase in the carbon diffusion coefficient does not compensate this effect.

**Conclusions**

The kinetic fundamentals of the formation of nanowires along the "vapor- liquid-solid" evolution passway are considered. It was shown that there are two basic stages: nucleation inside a binary nanodroplet and the growth of a nanowire, whose radius practically coincides with the nanodroplet radius.

In particular, for the cluster inside the nanodroplet and, partially, on the substrate, the expression for heterogeneous nucleation rate of is obtained. The growth of a nanowire is investigated. It was shown that the growth rate is directly proportional to the value of diffusion coefficient and inversely proportional to the nanodroplet radius. The prediction was made that the beginning of a nanowire has a porous structure on the length about 1nm. Useful limitations have been obtained on minimal sizes of nanodroplets. The growth rate of a tubular nanowire is investigated.

For a free molecular approximation estimation for the specific densities of fluxes, which are necessary for creating a steady - state supersaturated solution in nanodroplet, has been made. It is shown that release of latent heat of phase transition affects of growth of NW.

The contribution of surface diffusion of substance A is important only for small fluxes from a gas phase; the influence of surface diffusion on growth of NW was carefully considered in [1-3].

Finally it should be mentioned that many nanomaterials can approximately be considered as nanowires and, thereby, we can to use the obtained results for their theoretical and engineering analysis.

# References

1. V.G. Dubrovskii, and N.V. Sibirev, Rhys. Rev. E**70**, 031604 (2004).

2. N.V. Sibirev, I.P. Soshnikov, V.G. Dubrovskii, and E. Arshansky, Technical Physics Letters **32**, 292 (2006).

3. V.G. Dubrovskii, G.E. Cirlin, I.P. Soshnikov, A.A. Tonkikh, N.V. Sibirev, Yu.B. Samsonenko, and V.M. Ustinov, Phys. Rev. B**71**, 205325 (2005).

4. F.N. Borovik, and S.P. Fisenko, Technical Physics Letters **33**, 151 (2007).

5. S.P. Fisenko, B.N. Bazylev, and H. Wuerz, Journal of Engineering Physics and Thermophysics **76**, 743 (2003).

6. L.V. Landau, and E.M. Lifshitz, *Course of Theoretical Physics*, Vol. 10: *Physical Kinetics* (Pergamon, 1981).

7. V.P. Krainov, *Qualitative Methods in Physical Kinetics and Hydrodynamics* (American Institute of Physics, New York, 1992).

8. V.V. Chesnokov, R.A. Buyanov, and A.D. Afanas'ev, Izvestija of Siberian Branch AS USSR, ser. Chem. 1982. **2,** 60 (1982) (in Russian).

9. J.S. Rowlinson, and B. Widom, *Molecular Theory of Capillarity* (Clarendon, Oxford, 1982).

10. J.I. Frenkel, *Kinetic Theory of Liquids* (Dover, New York, 1955).

11. V.P. Scripov, *Metastable Liquids* (John Wiley and Sons, 1974).

12. S.P. Fisenko, and G. Wilemski, Phys. Rev. E**70**, 056119 (2004).

13. B.S. Kukharev, A.N. Rogozhnikov, S.P. Fisenko, and S.I. Shabunya, Journal of Engineering Physics and Thermophysics **65**, 804 (1993).

14. V.L. Kuznetsov, A.N. Usoltseva, A.L. Chuvilin, E.D. Obraztsova, and J.-M. Bonard Phys. Rev. B **64,** 235401 (2001).

15. M. Chhowalla, K. B. K. Teo, C. Ducati, et al, Journal of Applied Physics **90**, 5308 (2001).

16. H. Kanzow, and A. Ding, Phys. Rev. B**60**, 11180 (1999).

17. V.V. Chesnokov, R.A. Buyanov, and A.D. Afanasiev, Kinetics and Catalysis **24**, 1251 (1983) (in Russian).

18. V.V. Chesnokov, and R.A. Buyanov, Russian Chemical Reviews **69**, 623 (2000).

19. C. Klinke, J.-M. Bonard, and K. Kern, Phys. Rev. B**71**, 035403 (2005).

20. F. Glas and J.C. Harmand, Phys. Rev. B**73**, 155320 (2006).

21. D.J. Siegel, and J.C. Hamilton, Phys. Rev. B**68**, 094105 (2003).